\documentclass[sigconf]{acmart}

\usepackage{booktabs} 
\usepackage{todonotes}
\usepackage{amssymb}
\usepackage{listings} 
\usepackage{url}
\usepackage{comment}

\graphicspath{{./images/}}

\definecolor{colBPjs}{RGB}{0,100,100}
\definecolor{colBPjs2}{RGB}{0,100,200}
\definecolor{colJs}{HTML}{888800}
\definecolor{colComment}{RGB}{0,100,0}

\lstdefinelanguage{BPjs}{
  morekeywords={bp},
  sensitive=true
}
\lstdefinelanguage{text}{
  escapechar=\|,
  basicstyle=\ttfamily\footnotesize,   
  numbers=left,                   
  numberstyle=\footnotesize,          
  stepnumber=1,                   
  numbersep=5pt,                  
  showspaces=false,               
  showstringspaces=false,         
  showtabs=false,                 
  frame=single,                   
  tabsize=2,                      
  captionpos=b,                   
  breaklines=true,                
  breakatwhitespace=false,        
  numberstyle=\tiny\color{cyan},     
  emphstyle=\color{colJs},
  keywordstyle={\color{blue}},
  commentstyle=\color{colComment},   
}

\newcommand\nhcode[1]{\texttt{#1}}
\newcommand\code[1]{\texttt{#1}}
\newcommand\placeholder[1]{\todo[inline=true, color=yellow]{Placeholder: #1}}

\setcopyright{rightsretained}

\acmConference[MODELS'18]{ACM\\IEEE International Conference on Model Driven Engineering Languages and Systems}{October 2018}{Copenhagen, Denmark}
\acmYear{2018}
\copyrightyear{2018}

\begin{document}
\title{BPjs --- a framework for modeling reactive systems using a scripting language and BP}

\author{Michael Bar-Sinai}
\orcid{0000--0002--0153--8465}
\affiliation{
  \institution{Ben-Gurion University of the Negev}
}
\email{barsinam@cs.bgu.ac.il}

\author{Gera Weiss}
\affiliation{%
  \institution{Ben-Gurion University of the Negev}
}
\email{geraw@cs.bgu.ac.il}

\author{Reut Shmuel}
\affiliation{
  \institution{Ben-Gurion University of the Negev}
}
\email{reutsh@post.bgu.ac.il}



\begin{abstract}
We describe some progress towards a new common framework for model driven engineering, based on behavioral programming. The tool we have developed unifies almost all of the work done in behavioral programming so far, under a common set of interfaces. Its architecture supports pluggable event selection strategies, which can make models more intuitive and compact. Program state space can be traversed using various algorithms, such as DFS and A*. Furthermore, program state is represented in a way that enables scanning a state space using parallel and distributed algorithms. Executable models created with this tool can be directly embedded in Java applications, enabling a model-first approach to system engineering, where initially a model is created and verified, and then a working application is gradually built around the model. The model itself consists of a collection of small scripts written in JavaScript (hence ``BPjs''). Using a variety of case-studies, this paper shows how the combination of a lenient programming language with formal model analysis tools creates an efficient way of developing robust complex systems. Additionally, as we learned from an experimental course we ran, the usage of JavaScript make practitioners more amenable to using this system and, thus, model checking and model driven engineering. In addition to providing infrastructure for development and case-studies in behavioral programming, the tool is designed to serve as a common platform for research and innovation in behavioral programming and in model driven engineering in general.
\end{abstract}

%
%



\maketitle

\section{Introduction}

Behavioral Programming (BP) is a recently developed programming and modeling paradigm~\cite{harel2010programming}. Its inherent concurrent and modular nature offers a  dynamic approach for the design and creation of modern software systems. Moreover, when written according to simple guidelines, behavioral programs become executable models, which can be formally analyzed and verified. Thus, BP creates new opportunities for model driven engineering, and for software engineering in general.

Since its introduction in 2010, various libraries supporting BP have been developed~\cite{bpjecoop,harel2014scaling,Weinstock15,wwmerlang}. However, as each library was developed for a different purpose, they differ on various aspects of BP, which often makes them semantically incompatible with each other. This fractured landscape hinders BP-related research and dissemination, as it makes sharing and corroborating ideas harder, and code reuse prohibitively cumbersome.

This paper presents an attempt to support the existing body of work done in BP under consistent, unified semantics. We propose a generalized version of BP as a common denominator for the existing work and define ways it can be refined to support existing work. Additionally, we propose a communication protocol between BP models (b-programs) and  traditional software systems, which allows embedding b-programs as a component in heterogeneous systems.

We present BPjs: a software tool based on our proposed definitions. BPjs can analyze and execute behavioral programs, either standalone or embedded in traditional systems. This, as explained in the paper, allows for methodologies that combine a layer of formally verified components with layers of software that are certified only by testing.

The generalized BP presented in this paper, as implemented by BPjs, can serve as a common foundation for research in BP and in system engineering. In particular, it creates an easy and practical way of embedding models written using BP in regular systems written in Java. This opens new opportunities for model-driven engineering, in research, industry, and, as we have demonstrated during the 2018 fall semester, in the classroom.

The rest of this paper is organized as follows: Section~\ref{sec:bp-intro} provides a brief introduction to behavioral programming. Section~\ref{sec:bpjs} introduces BPjs, its design, and how it can be used for execution and verification of b-programs. Section~\ref{sec:sample_programs} introduces and briefly discusses four sample applications written in BPjs. Section~\ref{sec:evaluating_bpjs} discusses our experience using BPjs in various contexts, including research and teaching. Section~\ref{sec:related_work} discusses how the work presented in this paper relates to other works in BP, modeling, and verification. Section~\ref{sec:discussions} discusses various aspects of BPjs, ranging from software engineering in the context of BP to BPjs-specific issues. Finally, Section~\ref{sec:summary} concludes.

The code used in this paper is available on-line, either in~\cite{bpjs-code-appendix} or as part of the BPjs code repository. BPjs itself is open-sourced, and available on GitHub and Maven Central\footnote{See BPjs' project site at https://github.com/bthink-BGU/bpjs}.

\section{Behavioral Programming 101}
\label{sec:bp-intro}
Behavioral Programming (BP)~\cite{bp:site}, a variant of scenario-based programming, was introduced by Harel, Marron and Weiss in 2010~\cite{harel2010programming}. Under BP, programs are composed of threads of behavior, called b-threads. B-threads run in parallel, coordinating a sequence of events via a synchronized protocol, as follows: during program execution, when a b-thread wants to synchronize with its peers, it submits a statement to a central event arbiter, and then blocks. The statement that each b-thread submits declares which events it requests to be selected, which events it waits for (but does not request), and which events it would like to block. When all b-threads have submitted their statements, the arbiter selects a single event that was requested and not blocked. It then unblocks the b-threads that requested or waited for that event. The rest of the b-threads remain at their state, until an event they requested or waited for is selected.

This mechanism allows each b-threads to independently take care for a single scenario based requirement such as forbidding the scenario: "\code{TurnACOff}; Less than 1 minute passes; \code{TurnACOn}". The corresponding b-thread can, for example, wait for the event \code{TurnACOff} and then block the \code{TurnAcOn} event for the next minute (e.g., by counting 60 \code{SecondTick} events).


\section{Introducing BPjs} 
\label{sec:bpjs}
The tool we present in this paper is called BPjs. It is a platform aiming to support almost all\footnote{The only exception to this rule is the distributed event selection supported by BPC\cite{harel2014scaling}} the body of work created so far in behavioral programming under one roof. To this end, BPjs defines a generalized version of BP with well defined extension points and external interfaces. Extensions that use these interfaces can aid future work in BP, both by framing internal BP topics (``what are useful event selection strategies'') and external ones (``how can a BP subsystem be embedded in another system''). Thus, BPjs can serve as a common platform for researching and disseminating ideas in BP\@. Such platform was hitherto unavailable.

BPjs allows b-programs to be embedded in existing software systems. Sending data from a host application to the b-program is done by enqueuing events to the b-program's external event queue. Sending data from the b-program to the host is done via a listener interface, which informs the host which event was selected, as well as other program life-cycle events. A super-step based mechanism similar to the one proposed in~\cite{harel2014scaling} takes care for embedding the events withing the run of the program in a systematic way.

BPjs' generalized definition of BP standardizes two aspects that vary between former implementations: the algorithm that selects events, and the type of the events themselves. The event selection algorithm is abstracted under the concept of \emph{event selection strategy}. Given a b-program synchronization point, an event selection strategy calculates which events in it are selectable, which one of them is to be selected for the current calculation, and how this selection affects incoming events from the environment. Events in BPjs are standardized: they have a name (string), and optional data, whose type is unrestricted. Subsection~\ref{subs:event-selection-strategies} expands on event selection strategies in BPjs. BPC~\cite{harel2014scaling} also offered a generalized form of BP --- see Section~\ref{sec:related_work} for a short comparison between the two frameworks.

To the best of our knowledge, with the exception of distributed execution, BPjs can support all use-cases implemented by former research in BP\@. Porting those tools and algorithms to BPjs is left to future work, and to the community.

Next, we describe the static structure of BPjs, and then look at how it can be used to execute and verify b-programs. These descriptions are all high-level ones. For low level technical descriptions, we refer the reader to the project's website~\cite{bpjs:site}.

BPjs' design is informed by our experience building BP libraries for Java~\cite{harel2010programming}, C~\cite{harel2012non}, JavaScript~\cite{ashrov2015use}, Blockly~\cite{marron2012decentralized}, and other languages. It differs from previous BP code in that it is designed to serve as a common platform, rather than to fit specific research. Where previous code makes ad-hoc assumptions about various BP aspects, such as the type of events, or the way events are being selected, BPjs uses abstraction or generalization to support that specific case within its general framework. One abstraction example is our design decision to use the Strategy design pattern for selecting events --- previous code used hard-coded strategies (except for BPC~\cite{harel2014scaling}).

BPjs is implemented as a Java library that runs code written in JavaScript. It uses Mozilla Rhino\footnote{An open source JavaScript engine written and maintained by the Mozilla foundation. Rhino is available at https://developer.mozilla.org/en-US/docs/Mozilla/Projects/Rhino} JavaScript engine to execute regular JavaScript code, and custom code for handling synchronization calls. BPjs supports aforementioned extension and generalization using common design patterns, such as Strategy\cite{GoFBook} and Observer\cite{GoFBook}.

\begin{figure}
\centering
\includegraphics[width=3.5in]{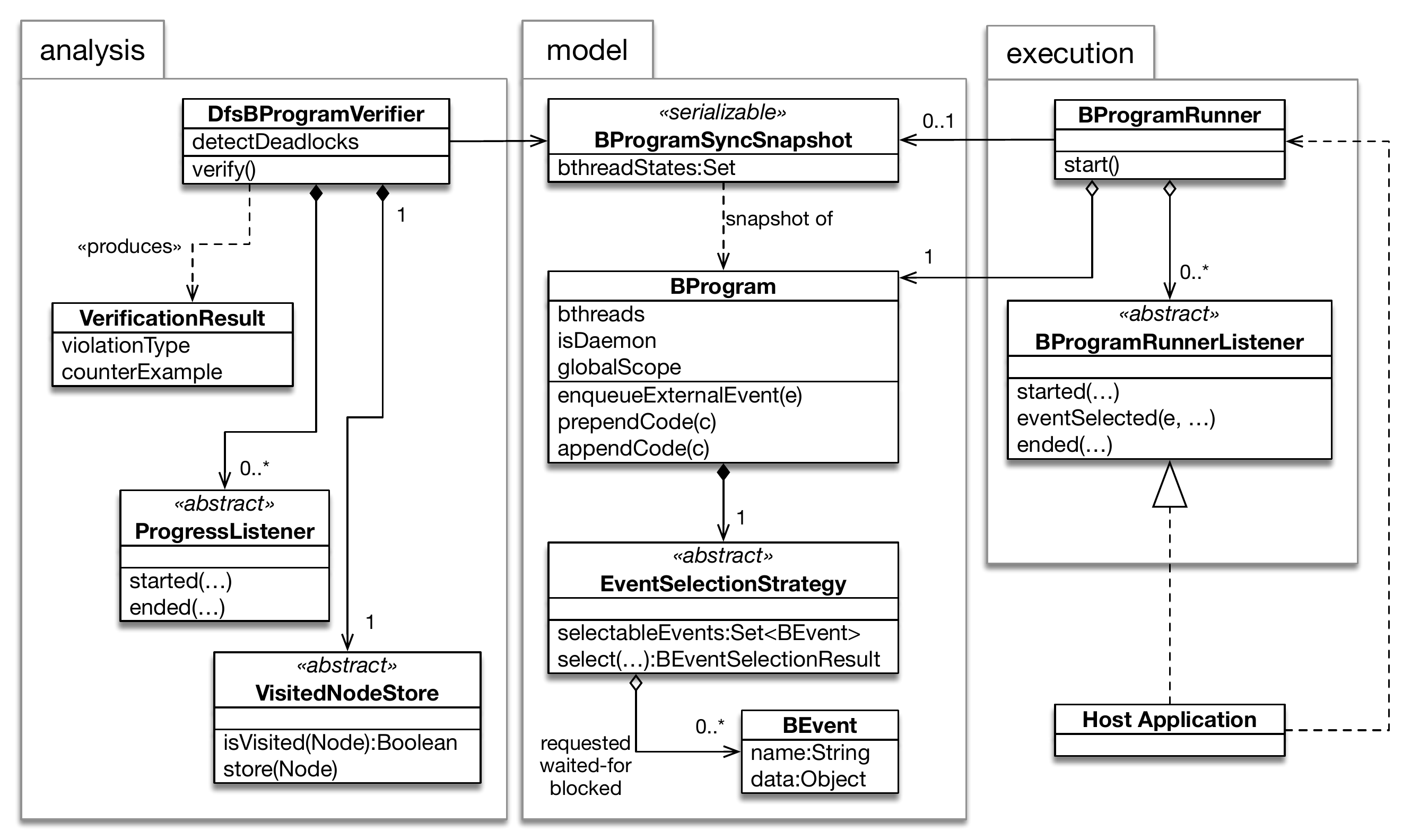}
\caption{Main classes of the BPjs platform (some details omitted for brevity). BPjs design maintains a strict division between model, execution and analysis. Classes in the \code{model} package are used to describe a b-program, but not to execute or analyze it. Model execution is done using classes in the \code{execution} package. The \code{analysis} package is used to verify and otherwise analyze a b-program.}
\label{fig:bpjs-main-classes}
\end{figure}

The main classes of BPjs are listed in \figurename{~\ref{fig:bpjs-main-classes}}. The \code{BProgram} class, which models a single b-program, is responsible for maintaining the state of a b-program: b-threads, global scope (presumably used only for read-only data), and an external event queue. A \code{BProgram} instance has a single \code{EventSelectionStrategy} object, which it uses for event selection during synchronization points. 

Perhaps surprisingly, the \code{BProgram} class does not have a \code{run} method. Nor does it have a \code{verify} one. This is because a b-program is somewhere on the continuum between a runnable program and a model; exactly where, is up to the b-program developers. In order to run a \code{BProgram} instance, we pass it to a \code{BProgramRunner}. In order to verify it, we pass it to a \code{DfsBProgramVerifier}, along with a formal specification and some additional information (see Sub Section~\ref{sub:verification} below). In this sense we say that a \code{BProgram} is an executable model --- we can just run it by advancing the states sequentially and we can also analyze it by traversing the states back and forth.

In addition to synchronizing, b-threads can call \emph{assert}, passing it a boolean expression as a parameter. If said expression evaluates to \code{false}, the b-program is considered in violation with its requirements. For added convenience, assertions also accept a string parameter, which allows programmers to explain what went wrong in a human-readable form. Assertions are common in programming languages, but incorporating them into BP is an addition introduced by BPjs.

\subsection{Program Execution} 
\label{sub:program_execution}

\begin{figure}
\centering
\includegraphics[width=3in]{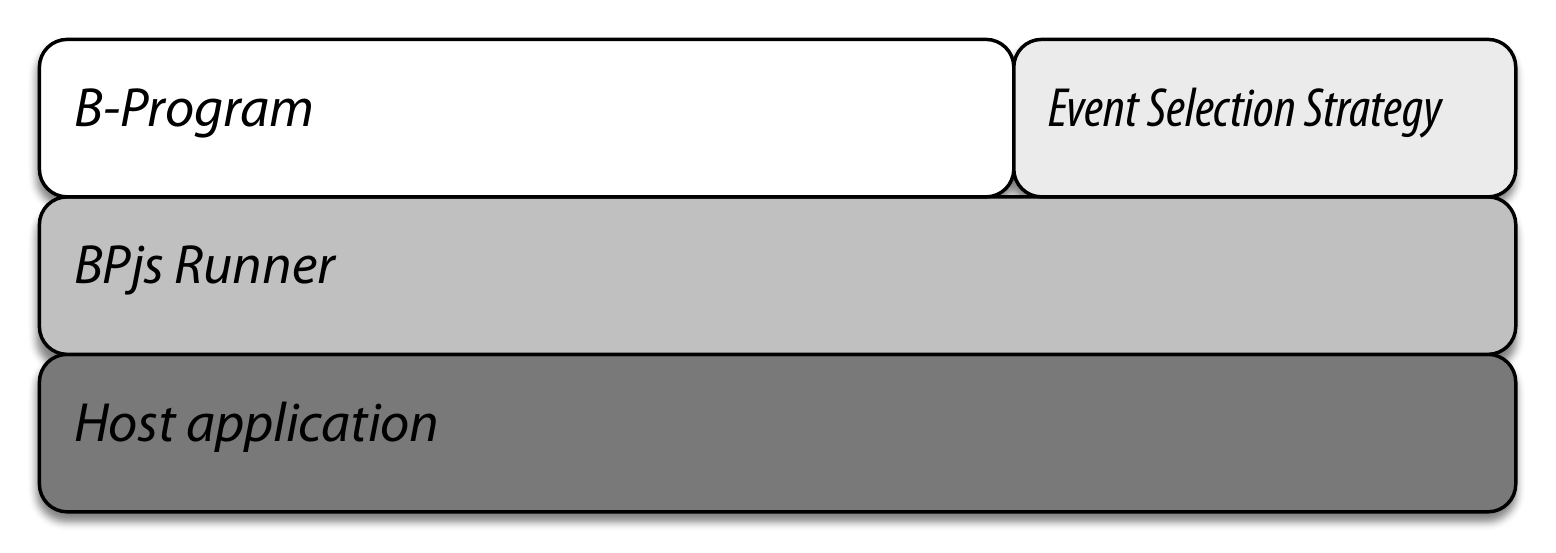}
\caption{BPjs program stack during b-program execution. The b-program and its event selection strategy run on top of a b-program runner, which handles program execution and communication with the host application. The host application reads data from the b-program by listening to the events it selects. It sends data to the b-program by placing events in its external event queue.}
\label{fig:bpjs-runtime-stack}
\end{figure}

The structure of an application running a b-program using BPjs is shown in \figurename{~\ref{fig:bpjs-runtime-stack}}. The b-program runs atop a b-program runner. Runners take a single b-program, and execute its setup code. Then, iteratively, they collect b-thread synchronization statements, use the b-program's event selection strategy to select an event that is requested and not blocked, and advance b-threads who requested or waited for said event to their next synchronization point. Additionally, runners maintain a list of b-program listeners, which is the protocol introduced by BPjs for monitoring a running b-program.

A b-program runner can be embedded in a host application. A host application can send data to the b-program by putting events in its external event queue. It can monitor the b-program by registering a b-program listener with the b-program runner.

For example, consider a bath controller whose logic is implemented using BPjs. When an \code{add\_hot\_water} event is selected, the host application is informed via the listener interface, and instructs its actuators to physically open the water stream. If its sensors detect that the bath is about to overflow, the host application enqueues an \code{overflow\_imminent} event into the b-program's event queue, so the b-program can respond. Section~\ref{sec:sample_programs} contains more examples.

Failed assertions cause an executing b-program to terminate. The host application is informed of the failed assertion, and may decide to shut down as well, drop into a ``safe mode'', or even restart the b-program under different settings, which may prevent the problem.

\subsection{Verification} 
\label{sub:verification}
\begin{figure}
\centering
\includegraphics[width=3in]{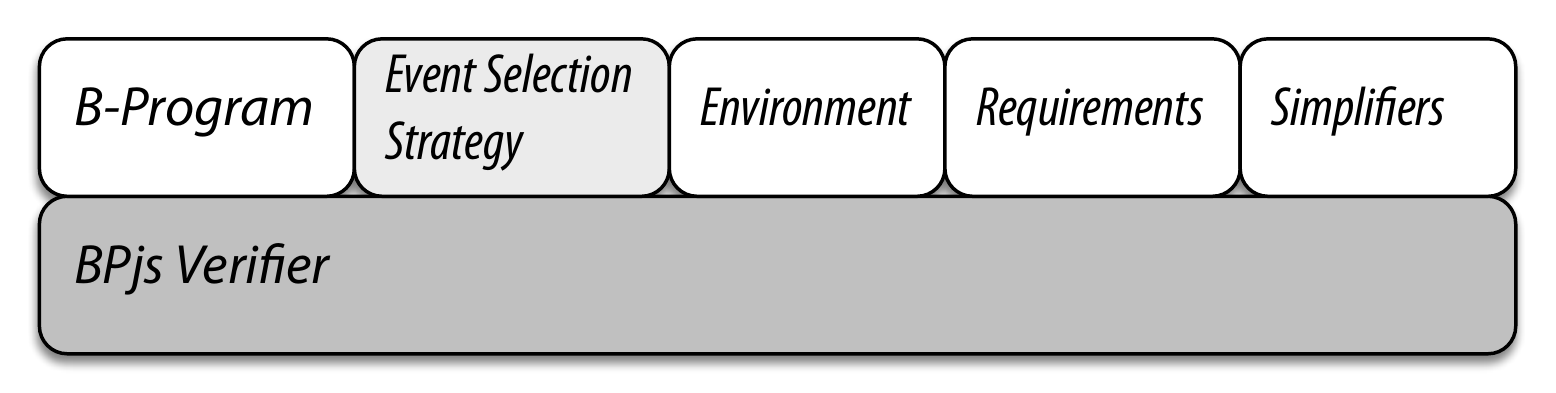}
\caption{BPjs program stack during b-program verification. The b-program and its event selection strategy are analyzed by a BPjs verifier. Additional BP code (white rectangles) models the system's formal specification, and environment. In some cases the search space can be reduced by adding b-threads modeling assumptions or specific domain knowledge.}
\label{fig:bpjs-analysis-stack}
\end{figure}

The structure of an application analyzing a b-program is shown in \figurename{~\ref{fig:bpjs-analysis-stack}}. The b-program and event strategy are directly analyzed, with no modification or transformation. Additional b-threads are added to the analyzed model in order to simulate the system's environment, and to detect requirement violations. Other b-threads can be added to limit the verification search space, which is typically rather large (as is the normal case in model checking). Sub Section~\ref{sub:sample-prog:mazes} shows an example of such a b-thread.

During verification, a model-checker traverses the system's state graph. When it discovers a new state, it checks that state for violations. There are two types of violations a verifier can detect: violating states, denoted by a b-thread that has a failed assertion, and deadlocks, where all requested events are blocked. The event sequence leading to the invalid state is used as a counter-example. Such sequences can later be used to fix a b-program, either manually, or automatically by an algorithm~\cite{harel2012non}.

\begin{figure}
\centering
\includegraphics[width=3in]{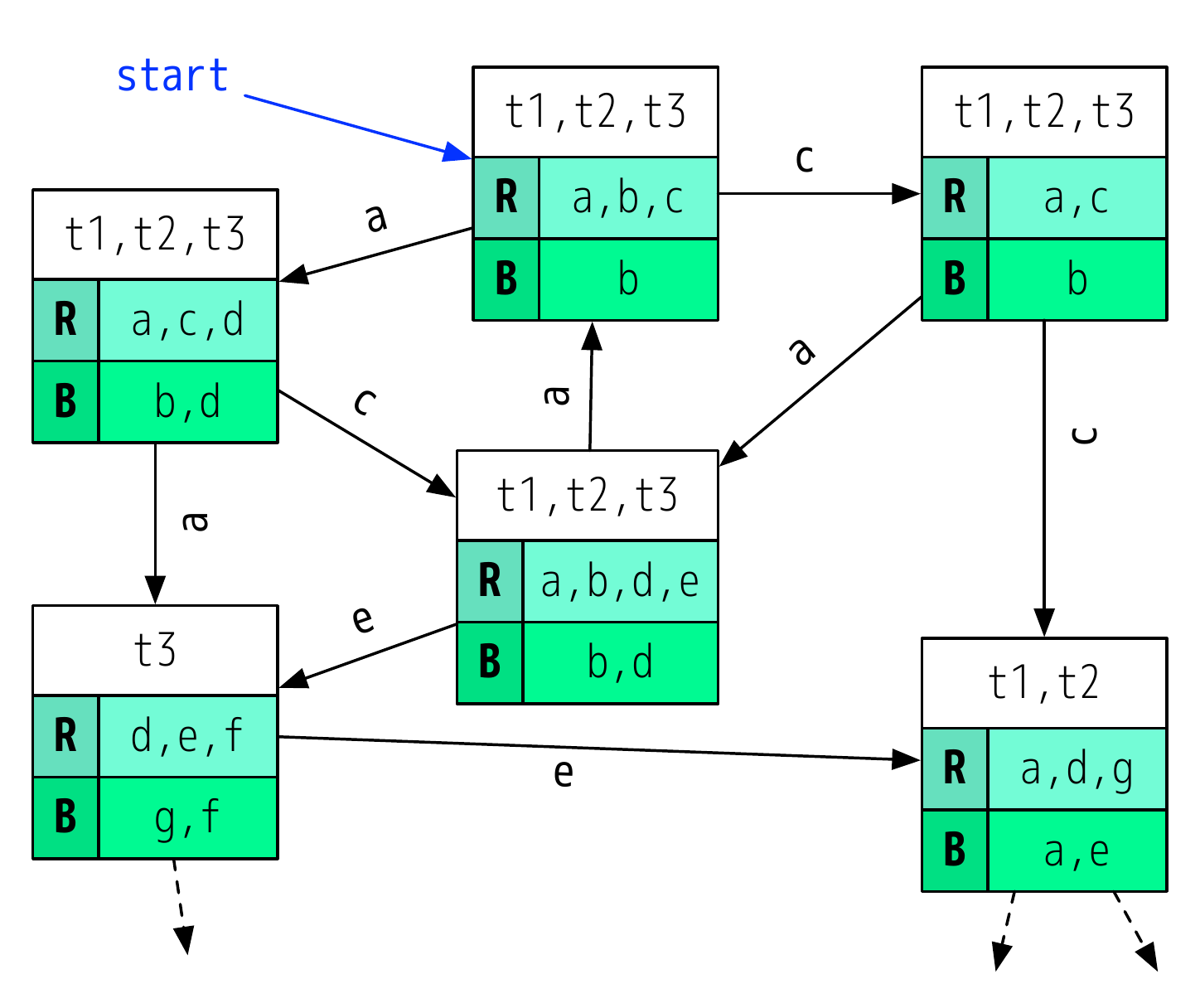}
\caption{State graph of a sample b-program. Each square represents a single program state. Each state contains b-threads participating in the b-program (top row), requested events (row $R$) and blocked events (row $B$). The program transitions between states by selecting events which are requested and not blocked (as marked on the edges). The graph is of a general shape --- we cannot assume the absence of cycles, bi-partitioning, or other simplifying properties.}
\label{fig:state-graph}
\end{figure}

In BP, a \emph{Program state} is defined as the states of all b-threads participating in a b-program, immediately after all of them have reached a synchronization point, and are thus paused, waiting for an event to be selected. The state of a single b-thread is defined as its stack, heap, and program counter. This definition of program state differs from the ``classic'' model checking definition, which considers each possible memory state to be a separate program state. However, BP semantics ensure that, as long as all inter-b-thread communication is done via events, stepping between synchronization points can be considered atomic~\cite{bpmc}. This allows BPjs program analyzers to look at a significantly smaller number of states, compared to, e.g.\ Java Pathfinder~\cite{Havelund:JavaPathfinder}. \figurename{~\ref{fig:state-graph}} shows an example b-program state graph.
Currently, BPjs uses depth-first search to traverse b-program state graphs. In~\cite{Weinstock15}, Weinstock used a very early version of BPjs to traverse a state space using A*. Concurrent and distributed state exploration are other feasible approaches. BPjs enables them by storing program state in a serializable form. Implementing such analyzers is left to future work.

The assert statement and the deadlock detector, combined with observer b-threads, allow specification (and hence, verification) of all safety properties of a b-program. However, they do not allow for specification of liveness properties. For example, in Subsection~\ref{sub:sample-prog:dining_philosophers} below, we cannot say with simple assertion that a philosopher eats infinitely often. To this end, BPjs allows for more general verification procedures. Specifically, one can extend \code{DfsBProgramVerifier} to also look for violating cycles in the state graph. For example, it is possible to specify that certain states are `hot' using the \code{assert} statement. Then, the generalized verifier may look for `hot cycles' --- reachable cycles whom states are all hot. This approach, which is similar in nature to the one taken by Live Sequence Charts (LSC)~\cite{Marron12:LSCRef}, allows for verification of all $\omega$-regular properties and more.

\section{Sample Programs} 
\label{sec:sample_programs}
This section illustrates a few aspects of creating systems with BPjs, using example programs. The full code, and a more detailed analysis of these examples, are available in BPjs' on-line documentation~\cite{bpjs:site}. Here we limit ourselves to a high-level discussion focusing on specific aspects of interest.

\subsection{Hot/Cold Bath} 
\label{sub:sample-prog:hot_cold_bath}

The hot/cold example, first introduced in~\cite{harel2010programming}, is the ``hello world'' program of Behavioral Programming, used as an example in numerous papers and tools to demonstrate how a basic b-program is written using a specific BP library. This program models a computerized bath controller, tasked with filling a bath with six parts water: three cold, and three hot. Said controller defines two events: \code{COLD}, and \code{HOT}. When the controller selects \code{COLD}, one part of cold water is added to the bath; when it selects \code{HOT}, one part of hot water are added to it.

Listing~\ref{lst:hotcold-basic} shows an implementation of such controller. Looking at its code, we can see that it is a regular JavaScript program. All BP-related calls are done by calling methods on the \nhcode{bp} object, which is defined globally. First, \nhcode{bp} is used to register two b-threads, \code{add-hot} and \code{add-cold}. B-threads in BPjs are regular JavaScript functions, labeled with a name. Then, when the regular JavaScript pass is over and the two b-threads have been registered, BPjs begins the b-program execution, where b-threads synchronize and events are selected.

Synchronization is done by invoking \nhcode{bp.sync} with a synchronization statement. BPjs allows passing partial statements, that is, if a b-thread does not wait for any events at a given point, the \nhcode{waitFor} field of the statement can be omitted. The events in this listing, \code{HOT} and \code{COLD}, are defined in the host application, and are programmatically added to the b-program's scope before it runs. This pattern is often used when an external system (in this case, the faucet actuators) has to respond to an event selected by a b-program.

\begin{lstlisting}[
  float,
  label={lst:hotcold-basic},
  mathescape=true,
  caption={A controller for filling a bath with three parts cold water, and three parts hot water. The order in which the water parts are added is left to the discretion of the event selection strategy.}
]
bp.registerBThread("add-hot", function(){
  bp.sync({request:HOT});
  bp.sync({request:HOT});
  bp.sync({request:HOT});
});
bp.registerBThread("add-cold", function(){
  bp.sync({request:COLD});
  bp.sync({request:COLD});
  bp.sync({request:COLD});
});
\end{lstlisting}

The code in Listing~\ref{lst:hotcold-basic} allows the bath to become too hot or too cold while being filled (for example, when \code{add-hot} runs to completion while \code{add-cold} remains blocked at its first synchronization point). To prevent these unbalanced scenarios, we can add an additional b-thread, such as the one in Listing~\ref{lst:hotcold-balancer}, to the bath controller b-program. The \code{control-temp} b-thread ensures the bath water temperature is never too hot, by blocking the addition of each part of hot water until a part of cold water is added. It is interesting to note that this b-thread can be added and removed without affecting the other b-threads, thus creating a modular functionality for the bath controller.

\begin{lstlisting}[
  float,
  label={lst:hotcold-balancer},
  caption={A b-thread that maintains safe water temperature by ensuring cold water are added before hot water are.}
]
bp.registerBThread("control-temp", function() {
  while ( true ) {
    bp.sync({waitFor:COLD, block:HOT});
    bp.sync({waitFor:HOT, block:COLD});
  }
});
\end{lstlisting}

\subsection{Dining Philosophers} 
\label{sub:sample-prog:dining_philosophers}
\begin{figure}
\centering
\includegraphics[width=2.5in]{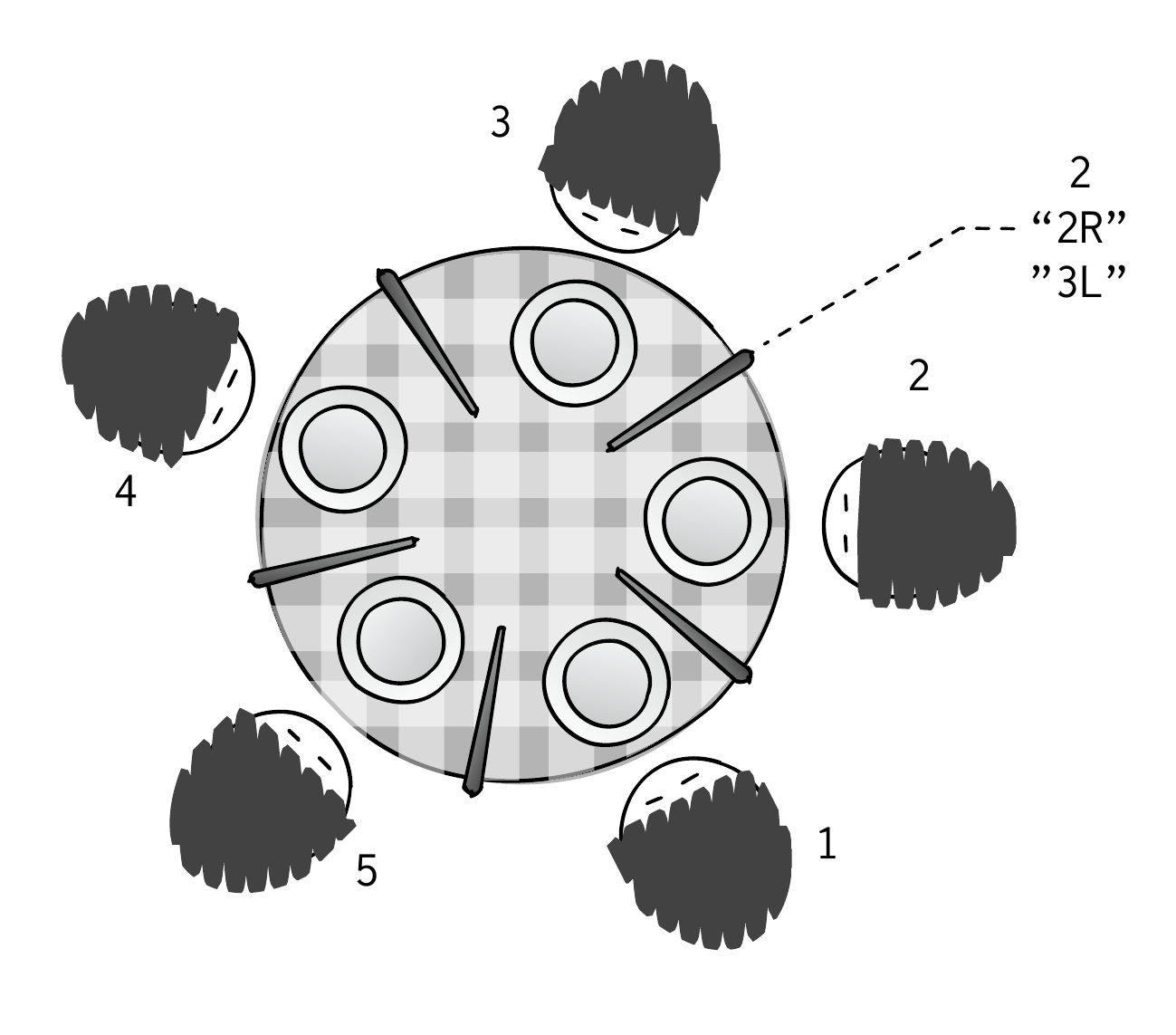}
\caption{Dijkstra's Dining Philosophers problem. Adjacent philosophers share the chopstick between them. In order to eat, a philosopher has to pick up the sticks left and right to her. Since a stick can be used by at most a single philosopher at a time, this setting poses mutual exclusion challenges.}
\label{fig:dining-philosophers}
\end{figure}

The Dining Philosophers is a classic concurrent programming challenge\footnote{The Dining Philosophers problem was first presented by Edsger Dijkstra in 1965, as an exam exercise.}. A group of philosophers is dining at a round table. For utensils, they have chopsticks - a single chopstick between each two plates. In order to eat, each philosopher has to obtain both chopsticks adjacent to her. This setting poses a mutual exclusivity challenge, as a chopstick may only be held by a single philosopher at any given moment. This problem is illustrated in Figure~\ref{fig:dining-philosophers}.

In this sub section, we look at a re-implementation of the Dining Philosophers b-program from~\cite{bpmc}, using BPjs. As modeled, philosophers use a naive algorithm to synchronize: they pick the chopstick to their right, then the one to their left, eat, and finally release the sticks in reverse order. This is a simple algorithm, but it can reach a deadlock.

The code in Listing~\ref{lst:dp-philosopher} shows a function for adding a philosopher b-thread to a b-program. Functions like these can be viewed as parametrized b-thread templates, as they create different b-threads based on the parameters they are invoked with. B-thread templates can be used to reduce code duplication, or to generate a heterogeneous b-thread population, e.g. by randomizing template parameters according to a required distribution.

\begin{lstlisting}[
  float,
  label={lst:dp-philosopher},
  caption={A function for adding a philosopher to the dining philosophers b-program. A dining philosopher repeatedly attempts to pick the chopstick to her right, then the one to her left, and then releases them in reverse order. }
]
function addPhil(philNum) {
  bp.registerBThread("Phil"+philNum, function() {
    while (true) {
      // Request to pick the right stick
      bsync({
        request: bp.Event("Pick"+philNum+"R")
      });

      // Request to pick the left stick
      bsync({
        request: bp.Event("Pick"+philNum+"L")
      });

      // Request to release the left stick
      bsync({
        request: bp.Event("Rel"+philNum+"L")
      });

      // Request to release the right stick
      bsync({
        request: bp.Event("Rel"+philNum+"R")
      });
    }
  });
};
\end{lstlisting}

Enforcing restrictions on chopstick usage, such as mutual exclusivity, is done by chopstick b-threads\footnote{The code for the chopstick b-threads, as well as the full program, are available on-line at~\cite{bpjs:site}.}. Each chopstick is modeled by a single b-thread, that blocks events based on its current state. For example, after chopstick \#2 is picked up by philosopher \#2 (that is, after event \code{Pick2R} was selected), it prevents philosopher \#3 from picking it up, by blocking event \code{Pick3L}. This block is lifted after philosopher \#2 releases the chopstick (\code{Rel2R}).

Since we use b-thread template functions for adding philosophers and chopsticks to the model, the creation the actual simulation boils down to a single loop calling theses functions (see Listing~\ref{lst:dp-create}).

\begin{lstlisting}[
  float,
  label={lst:dp-create},
  mathescape=true,
  caption={Creating a dining philosophers model.}
]
for (var i=1; i<=PHILOSOPHER_COUNT; i++) {
  addStick(i);
  addPhil(i);
}
\end{lstlisting}

The dining philosophers b-program described here can serve both as a simulation program and as a model to be checked. For simulation purposes, this b-program is run, and its event log can be analyzed, e.g.\ to get statistics about stick wait times. For verification, the b-program is passed to a verifier for deadlock detection. In both cases, the exact same code is used --- no translation is necessary when transitioning between code execution and verification. In terms of model checking performance, using a 2.6 GHz laptop with 16GB ram, BPjs finds a deadlocking counterexample for 20 dining philosophers, in about 2 seconds. For 30 dining philosophers, finding a counter example takes slightly more than 4 seconds.

\subsection{Mazes and Model Checking} 
\label{sub:sample-prog:mazes}
\begin{figure}
\centering
\includegraphics[width=2.5in]{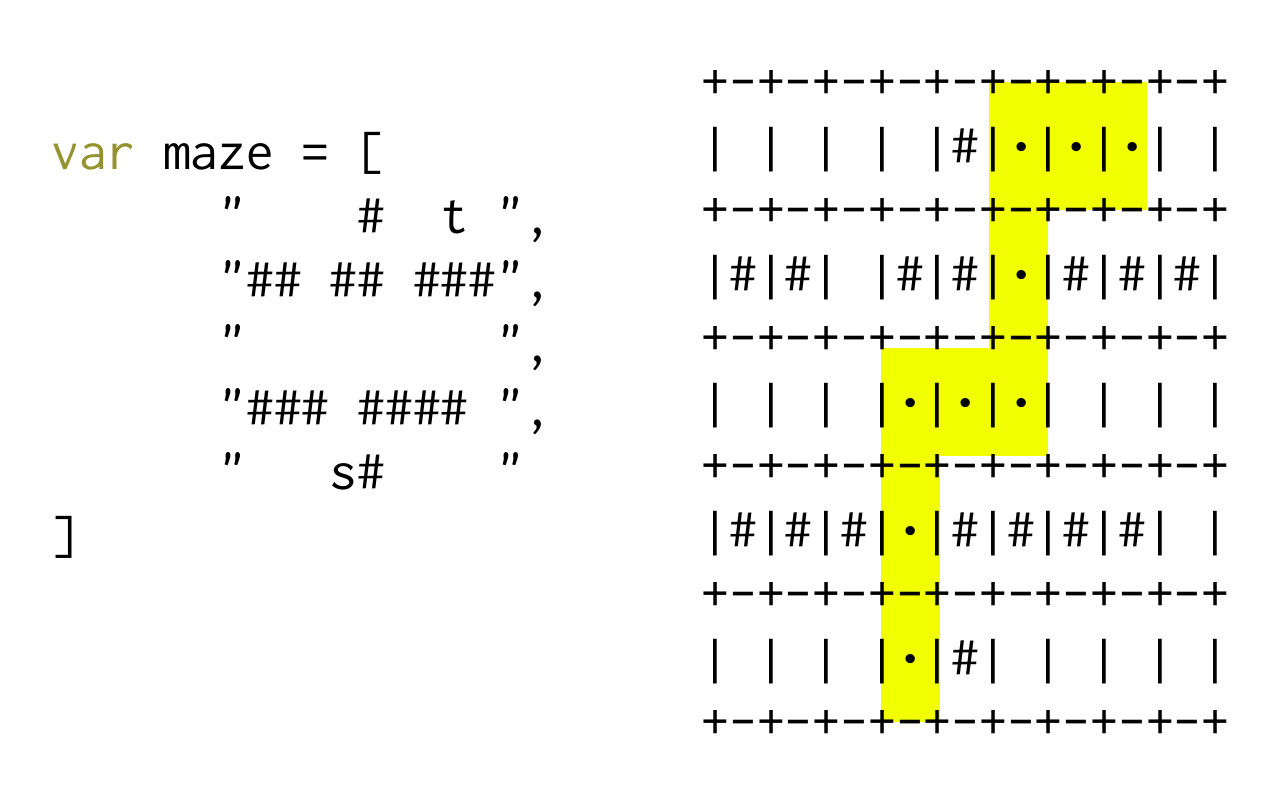}
\caption{A maze described using our maze-description DSL (left), and its solution, found by verifying the b-program that models it (right).}
\label{fig:maze}
\end{figure}

In this subsection, we apply techniques presented in~\cite{SemanticVariationsExe16}, using BPjs, for solving mazes. To this end, we will define a simple domain specific language (DSL) for describing mazes. In our DSL, mazes are created using ASCII drawings (see example at \figurename{~\ref{fig:maze}}). Each character in such drawing has specific semantics: spaces denote places a maze walker can walk into, \code{s} denotes the starting point for the walker, and \code{t} denotes a target cell. All other characters denote walls.

Parsing a maze written in this DSL amounts to traversing an ASCII drawing, and adding appropriate b-threads for each of its characters. Listing~\ref{lst:maze-cell} shows the b-thread template used to generate b-threads for space cells.

Under our model, space cells attract the maze walker when it enters a cell adjacent to them. To do this, they repeatedly wait for an entry event to any of their neighboring cells, and then request an entry event with their own coordinates. Note the parameters passed to \nhcode{waitFor}: these are not events; they are event \emph{sets}. An event set is a predicate, accepting an event and returning \code{true} iff that event is a member of the set. Event sets can be defined statically, like \code{anyEntrance}, or dynamically, by composing a new predicate function based on a set of parameters. This is the case in \code{adjacentCellEntries}, used in line 8.

\begin{lstlisting}[
  float,
  label={lst:maze-cell},
  mathescape=true,
  caption={A b-thread template for adding a space cell at \code{row}, \code{col}. Space cells wait for the maze walker to enter a cell adjacent to them, and then request that it enters them. If the event selection strategy chose another cell for the maze walker to enter, the entry request is removed.}
]   
var anyEntrance = bp.EventSet("AnyEntrance", function(e){
   return e.name.indexOf("Enter") === 0;
});

function addSpaceCell( col, row ) {
  bp.registerBThread("cell(c:"+col+" r:"+row+")",
    function() {
      while ( true ) {
        bsync({waitFor: adjacentCellEntries(col, row)});
        bsync({request: enterEvent(col, row),
               waitFor: anyEntrance});
      }
    }
  );
}
\end{lstlisting}

Target cells are regular space cells, with an additional trait: when the maze walker enters them, \code{TARGET\_FOUND\_EVENT} is selected. To achieve this, the parser adds an additional b-thread for each target cell. This b-thread waits for the relevant entry event, and then requests \code{TARGET\_FOUND\_EVENT}, while blocking all other events. 

Similarly, a start cell is a space cell with an additional behavior: when the maze program starts, it already has a maze walker in it. Thus, when handling a start cell, the maze parser generates an additional b-thread that requests an entry event with that cell's coordinates when the b-program starts. 

Parsing a maze description results in a b-program modeling the maze. Running the b-program will make the maze walker perform a valid random walk through the maze. It will start at the start cell, will not pass through walls, and if it will stumble on the target cell, \code{TARGET\_FOUND\_EVENT} will be fired. To solve a maze efficiently, we can use a verification.

Our formal requirement will be that \code{TARGET\_FOUND\_EVENT} is never fired. A b-thread modeling this requirement will \nhcode{watFor} this event, and then call \nhcode{bp.ASSERT(false)}, to indicate that a requirement has been violated. After adding this b-thread to the generated b-program, we can pass it to a b-program verifier. If the maze has a solution, the verifier will return a counter example, with the path to the target cells encoded in its event trace.

Verification of a maze b-program can take a long time and generate an arbitrarily long counter example. This is because of the random walk: while a counterexample will be found, the length of the path it uses to get to the target cell is unbounded. We can solve this issue by adding a b-thread that prevents a the maze walker from entering the same cell twice (see Listing~\ref{lst:mazes-once}). This thread turns the random walk into a walk that enters a new cell at every step, until it gets terminally stuck. The verification process examines all such possible walks, and returns one that ends up at the target cell. This is an example of a \emph{simplifier} b-thread, which reduces the search space required for verification. Note that adding this b-thread eliminates an infinite number of maze solutions: all paths from start to target that visit a cell more than once. However, for each eliminated solution, there exists a solution where each cell along the path is visited only once. This solution, which is a local minimum, remains in the search space. Thus, the verification is sound, even though it does not traverse the entire state space.

\begin{lstlisting}[
  float,
  label={lst:mazes-once},
  caption={A b-thread preventing a the maze walker from entering a cell twice. This b-thread during verification, this b-thread reduces the search space, while maintaining the soundness of the verification result.}
]
bp.registerBThread("onlyOnce", function(){
  var block = [];
  while (true) {
    var evt = bsync({waitFor: anyEntrance,
                       block: block});
    block.push(evt);
  }
});
\end{lstlisting}

\subsection{Tic Tac Toe} 
\label{sub:sample-prog:tic_tac_toe}
In this subsection, we use a TicTacToe program to demonstrate the usage of a priority-based event selection strategy, model execution and analysis, and environment simulation. The b-program presented here is borrowed, with modifications, from~\cite{bpjecoop}, where is was used to demonstrate the concept of aligning b-threads with system requirements.

Our Tic Tac Toe program plays against a human on a $3 \times 3$ board, using classic Tic Tac Toe rules. It is made of two groups of b-threads. The first group enforces the rules of the game, such as ``players take turns'' (using a b-thread similar to the temperature-regulating b-thread in Listing~\ref{lst:hotcold-balancer}), or ``a square can be marked only once'' (using a b-threaed similar to the simplifier thread in Listing~\ref{lst:mazes-once}). B-threads from this group are also responsible for detecting when a player wins, or if a game ends in a tie.

The second group forms the strategy of the computer player. It is composed of numerous b-threads that are responsible for defending against opponent attacks, and presenting initiative when possible. Listing~\ref{lst:ttt-strategy} shows two of these b-threads. Assume that \code{line} holds an array of three squares that forms a line on the board (i.e. a row, column, or diagonal). If player O have already placed $O$s in the first two squares, the b-thread \code{AddThirdO} suggests placing an $O$ in the remaining square, winning the game for player O. Similarly, if player X have already placed $X$s in the first two cells of that line, b-thread \code{PreventThirdX} requests putting an $O$ at the third square, in order to prevent the opponent from winning. 

The calls to \nhcode{bp.sync} where the aforementioned events are requested use an additional parameter - the priority of the request. Here, adding the third $O$ has a higher priority than preventing a third $X$. This is because the strategy prefers an immediate win over preventing a loss and continuing the game.

The priority semantics of the second \nhcode{bp.sync} parameter are casted by the event selection strategy used by the Tic Tac Toe b-program. From BPjs' point of view, the second parameter has no defined semantics --- it is a hint to the b-program's event selection strategy. It becomes a part of the calling b-thread's synchronization statement, and BPjs passes it to the event selection strategy as-is. This allows BPjs to support multiple event selection strategies in a fully pluggable way. 

\begin{lstlisting}[
float,
label={lst:ttt-strategy},
caption={Two b-threads that are part of the player strategy. \code{AddThirdO} is responsible for completing lines of two $O$s into a triplet (and, thus, winning the game). \code{PreventThirdX} is responsible for preventing the opponent from doing the same. The usage of requests with priorities (lines 5 and 12) ensures that, when required to choose between completing a row and blocking an opponent, the strategy will always prefer the former.}
]
bp.registerBThread("AddThirdO", function() {
  while (true) {
    bp.sync({waitFor:[O(line[0].x, line[0].y)]});
    bp.sync({waitFor:[O(line[1].x, line[1].y)]});
    bp.sync({request:[O(line[2].x, line[2].y)]}, 50);
  }
});
bp.registerBThread("PreventThirdX", function() {
  while (true) {
    bp.sync({waitFor:[X(line[0].x, line[0].y)]});
    bp.sync({waitFor:[X(line[1].x, line[1].y)]});
    bp.sync({request:[O(line[2].x, line[2].y)]}, 40);
  }
});
\end{lstlisting}

The Tic Tac Toe b-program can be used in two contexts: as a controller part in an interactive application, and as a model being verified. In a GUI context, a host GUI application runs the b-program internally. User clicks are translated to \code{X(x,y)} events and placed in the b-program's external event queue. The host application listens to events selected by its b-program, and updates its UI accordingly (e.g. drawing an $X$ in the top-right square when \code{X(2,0)} is selected). 

In a verification context, we use the same b-program, and add b-threads that model requirements. Listing~\ref{lst:ttt-x-never-win} shows a b-thread modeling the requirement ``X should never win''. Verifying our strategy against this requirement ensures that we have created a good Tic Tac Toe strategy\footnote{Tic Tac Toe does not have a strategy that guarantees victory. There exists a strategy which guarantees not losing the game, though.}.

\begin{lstlisting}[
float,
label={lst:ttt-x-never-win},
caption={A b-thread declaring that a state where X has won is illegal. This b-thread is a direct translation of the requirement ``X should never win''.}
]
bp.registerBThread("R1:XShouldNotWin", function(){ 
  bsync({waitFor:bp.Event(XWin)});
  bp.ASSERT(false, "X won.");
});
\end{lstlisting}

Having added a formal specification, we now need to validate the strategy against all possible opponents. Ostensibly, this is a daunting task. However due to the fact that, when faced with multiple options for advancing, the verifier examines all of them, we can perform the required verification by adding a single b-thread, shown in Listing~\ref{lst:ttt-simulated-player}. The \code{SimulatedOpponent} b-thread constantly requests placing $X$s in all squares on the board. The game rule b-threads enforce that those requests are only honored during X's turn, and for non-populated squares. Thus, for each of X's possible turns, the verifier tests the consequences of placing an $X$ in each of the available locations.

\begin{lstlisting}[
float,
label={lst:ttt-simulated-player},
caption={A b-thread that simulates all possible behaviors of player X during model checking. The fact that this b-thread requests all squares, combined with the model checker state scanning, results in the model checker traversing the program execution sub-tree for every legal choice of placing an X on the board. Requests for placing an X in illegal locations are blocked by the game rules b-threads.}
]
bp.registerBThread("SimulatedOpponent", function(){
  while (true) { 
    bsync({request:[X(0,0),...,X(2,2)]},10);
  }
}
\end{lstlisting}



\section{Evaluating BPjs} 
\label{sec:evaluating_bpjs}
BPjs is designed to support multiple BP variants, be used in different settings, and maintain a non-intimidating, developer-friendly nature. In this section we look at our experience using BPjs, and try to asses whether these goals were achieved.

In its current form, BPjs was first used in~\cite{SemanticVariationsExe16} to run specifications written in LSC~\cite{DH01a}. To that end, authors embedded BPjs in a command-line application. The application parsed LSC diagrams into b-programs written in JavaScript, and passed these programs to BPjs for execution. Internally, our group used BPjs to program a robot in Robocode~\cite{Robocode2004}, and in a desktop GUI application (the Tic Tac Toe example presented in this paper). Recently, a team of students and researchers embedded BPjs in a web server in order to implement a graphical, web-based rule engine. The system was developed during a 28 hours hackathon, and went on to win first prize\footnote{First place in HackBGU, https://bit.ly/2riUjB0}. Lastly, we are currently building a prototype of an on-board satellite control software based on BPjs. This project was funded by the Israel Innovation Authority, after a board of external experts examined BPjs and BP, and found them fitting for satellite control. Thus, we feel confident saying that BPjs can be used successfully in different settings.

In order to test how approachable BPjs is, we have used it as a teaching environment in an undergraduate course for computer science (CS) and information systems engineering (ISE) students. Students chose between implementing a project in BPjs or in an IoT environment. Out of 42 students, 20 chose to work with BPjs. Implemented projects included a web-based PacMan game, a Blockly based interface for behavioral programming, an implementation of strategies in computer games, and more. Students were instructed to first read the tutorials and only then approach the authors with questions regarding BPjs. Except for a single case where students implemented their own event selection strategy, the tutorials proved to be enough for students to implement their projects.

In the course feedback, a 3rd year CS student writes: ``the whole idea of this system is very interesting for me, because it is actually a different approach to problems than the approach we, as students in CS, are used to have. The way BP engine works \ldots might be strange in the beginning, but when I got into it - it looked really logical and obvious''. Another student writes: ``Using the decision engine based on request, wait-for, and block, was initially hard to understand, but after a few examples I was able to understand it and enjoy using it. I found this way of thinking to be interesting and challenging''. One student concluded his feedback saying: ``Finally, I can say that this system is revolutionary in the way it sees and solves problems, but at the same time really friendly to the user.'' 

At the end of the course, a student survey was conducted, yielding the following results:
\begin{center}
    \begin{tabular}{ | p{4.5cm} | p{2.5cm} |}
    \hline 
    \textbf{Question} & \textbf{Answer Average (0-5)} \\ \hline \hline
    The material was presented in a clear and organized manner. & 4.7 \\ \hline
    I found the material and the way it was presented interesting.  & 4.6 \\ \hline
    The use of the tool motivated me to be creative.  & 4.7 \\ \hline
    I found the material taught in class relevant. & 4.5 \\ \hline
    Overall satisfaction.  & 4.6 \\ \hline
    \end{tabular}
\end{center}


\section{Related Work} 
\label{sec:related_work}

The closest relative to BPjs is BPj. Introduced in~\cite{harel2010programming}, the paper that also introduced Behavioral Programming, BPj allows writing behavioral programs using Java. Unlike BPjs, BPj was never designed as a general framework. Over time, a few mutually incompatible versions emerged, as researchers adapted it to fit specific needs. For technical reasons, BPj does not support verification beyond Java 5. Like BPjs, BPj is an open source software. However, its infrastructure is not adjusted to modern open source standards: it uses an IDE-specific project structure, and was developed in a close repository with the source code available as a zip archive.

Our work on BPj informed many of our decision while working on BPjs. In particular, the decisions to use modular design, an IDE-agnostic project format, and an open, collaboration-friendly code repository.

BPC~\cite{harel2014scaling} is a framework for writing behavioral programs in C++. Like BPjs, it offers customizable events and event selection. BPC also supports time constraints, a feature not currently supported by BPjs. BPC supports limited verification, based on examining the state graph of each b-thread, and then composing and analyzing the program graph using an external tool. A key difference between BPC and BPjs is that BPC was not planned to be embedded in host applications, but rather to be run standalone. But, as demonstrated in Section~\ref{sec:evaluating_bpjs}, the ability to embed a b-program in a larger system opens many interesting use cases, for which BPC cannot be used. 

Both BPC and BPj use a single system thread per b-thread, which makes b-threads expensive. This is an issue, since under the BP paradigm, b-programs should be able to consist of many small b-threads. In BPjs, on the other hand, a single system thread can run multiple b-threads, which makes b-threads much cheaper. This is a technical difference, but it enables BPjs to run larger b-programs using less resources.

An obvious difference between BPjs, BPC, and BPj, is the language used for writing b-threads. While BPC and BPj use a programming language, complete with static typing and strict class definitions, BPjs uses a lenient scripting language. We claim that scripting languages are generally a better choice for BP, because the code for b-threads is normally quite short. Short programs don't require much code organization, and so the syntactic overhead imposed by programming languages in order to organize their code is non-beneficial for b-threads. Scripting languages allow writing cleaner b-thread code, as they skip a lot of boilerplate.

PlayEngine~\cite{HM03}, PlayGO~\cite{playgo}, and Scenario Tools~\cite{greenyer2017scenariotools} support BP by implementing Live sequence charts (LSC)~\cite{DH01a}. BPjs, as well as BPC and BPj, offer more generality, but lack the slick visual interface these tools have.

Looking more broadly at executable modeling languages, Executable UML\cite{mellor2002executable} models general programs using diagrams with executable semantics, and an action language. In~\cite{SemanticVariationsExe16} we have shown how BPjs can be used to define executable semantics of diagrammatic languages, and applied the proposed technique to UML's Sequence Diagrams. In this capacity, BPjs can provide a common meta-language for executable UML diagrams.

JavaPathfinder (JPF)~\cite{JPF} performs verification on directly on Java code. Like BPjs, JPF uses its own runtime mechanism to execute the analyzed program. Another similarity is that both JPF and BPjs are designed to be very modular. The main difference between BPjs and JPF is that BPjs assumes it is verifying a behavioral program, and thus can limit the states it examines to synchronization points only. JPF cannot make this assumption, and thus has to process a significantly larger amount of states.

\section{Discussions}
\label{sec:discussions}

This section briefly discusses options and questions raised by BPjs regarding Behavioral Programming, and software engineering in general. These discussions are not meant to be exhaustive, but rather to serve as starting points for a conversation.

\subsection{Event Selection Strategies}
\label{subs:event-selection-strategies}

Behavioral Programming requires that at every synchronization point, only events that are requested and not blocked can be selected. The decision which event should be selected is left for the implementation. Indeed, different BP implementations choose their events in different ways. The original implementation~\cite{harel2010programming} prioritized events according to the b-thread that requested them. Subsequent versions used randomized selection, among other strategies.

BPjs introduces a unified interface for event selection algorithms, which, for the best of our knowledge, supports all existing algorithms. This interface defines two methods: the first accepts the state of a b-program at a synchronization point, and returns a set of selectable events. The second method accepts a set of selectable events, and a b-program's state, and performs the actual selection. This former method is used during verification, while execution uses both. This separation allows event selection algorithms to use randomized event selection, when more than a single event is selectable.

In order to support all existing strategies, BPjs allows b-threads to pass a meta-data object to \nhcode{bp.sync}, in addition to the usual statement declaring which events are requested, waited-for, and blocked. The additional object may guide the event selection strategy's choice. BPjs itself includes four different event selection strategies, each of which was found useful, either by us or by prior work.

\begin{enumerate}
  \item \emph{Simple} Selects a random event that is requested and not blocked.
  \item \emph{Prioritized B-Threads} Maintains a priority for each b-thread. Selects a non-blocked event, requested by a b-thread with the highest priority who requested such event.
  \item \emph{Prioritized Synchronizations} B-threads may add a priority to their \nhcode{bp.sync} statements. Given a set of events that are requested and not blocked, this strategy will randomly select an event with the maximum priority. The Tic Tac Toe example presented in Sub Section~\ref{sub:sample-prog:tic_tac_toe} makes use of this strategy.
  \item \emph{Ordered Events} BPjs allows a b-thread to request multiple events by passing an event array to \nhcode{bp.sync}. Other strategies treat this array as a set of events; this strategy treats it as a list. Thus, for each b-thread, it will only consider selecting the first requested and not blocked event.
\end{enumerate}

Many other options for event selections spring to mind. For example, events can be selected using majority vote (calculated using \nhcode{request}s or \nhcode{waitFor}s). Algorithms can use planning, heuristics, or consult machine learning algorithms. 

Event selection algorithms can use analysis as a way of informing event selection decisions. The basic idea, similar to the one used by many chess playing programs, is to see what are the likely outcomes of selecting each of the selectable event at a given moment. To achieve this, an event selection strategy can preform a limited verification session, starting at the state in question, and selecting each possible event. The verification does not have to be exhaustive --- it may be enough to see what states are reachable within a limited amount of steps, and how likely is the program to actually reach them. Given this data, and a scoring algorithm for each reachable state, the event selection strategy will have valuable information to work with when deciding which event to select.

\paragraph*{Hierarchies} Event selection strategies can be ordered by hierarchy, according to the event sequences they allow. An event selection strategy $s_{child}$ is under event strategy $s_{parent}$ if all event sequences made possible by $s_{child}$ are also possible under $s_{parent}$. Thus, if a system was verified under $s_{parent}$, it is also valid under $s_{child}$.

The Simple event selection strategy, which selects an event which is ``requested and not blocked'' is at the top of this hierarchy, as it allows all event sequences. Thus, if a system was verified with the Simple event selection strategy, it will be valid with \emph{any} event selection strategy.

This type of event selection strategies present an interesting combination between forward execution and model checking, creating a smart runtime, as described above. Of course, as verification can take time, an event selection strategy may choose to use verification only when it has enough time (e.g. in a drone, with no obstacles close by), and use a faster event selection algorithm when event selection has to be done fast.

\subsection{Development Process of a Behavioral System} 
\label{sub:development_process_in_bpjs}

The process of developing a software system using BPjs differs from the regular process of developing a system, in that developers may directly verify the system being developed against a set formal requirements. Conveniently, the requirements and the system are described using the same language, tools, and formalisms.

A system developed in BPjs is composed of b-threads. Normally, each b-thread assumes one of the following roles:

\begin{enumerate}
  \item\emph{Model b-threads} These b-threads form the core of the system. The equivalents of source code in regular software development, these b-threads will ship as part of the production system, once it is ready.
  \item\emph{Requirement b-threads} These b-threads form the formal requirements of the system. They follow the progression of the system while making assertions about it. When such assertion fails, the system is considered to be in violation of the requirement. Depending on the verification done before shipping the system, requirement b-threads may or may not be part of the complete product. Requirement b-thread against which the b-program was validated can be removed from the system, as the developers can safely assume that none of the requirements they represent are violated. Requirement b-threads that were not validated against, can ship as part of the finished product, and halt the system in case a violation is made. This is a form of monitoring, or ``runtime verification''~\cite{runtimeVerification}.
  \item\emph{Environment b-threads} These b-threads are used during the verification process to simulate the environment the system interacts with (Listing~\ref{lst:ttt-simulated-player} is an example of such a b-thread). These b-threads are not part of the shipped product. In this capacity, they are similar to unit testing code in mainstream software development.
  \item\emph{Assumptions b-threads} These b-threads allow faster verification process, by narrowing the amount of execution possibilities. They are important, as verification of a non-trivial system can take a long time. However, they should be used with care, since they may eliminate cases that might occur during execution. Such b-threads should be written in a way that makes the scenarios they eliminate very clear. One example for these b-threads is the \code{onlyOnce} b-thread of the mazes example (Listing~\ref{lst:mazes-once}). This b-thread eliminates an infinite number of runs by preventing the maze walker from entering the same cell twice. However, said elimination does not invalidate the verification process, since for each path from \code{S} to \code{T} that contains repeated entries to cells, we can build an equivalent simple path, that also arrives at the target cell. Thus, if there's a path from \code{S} to \code{T} in the given maze, \code{onlyOnce} would not prevent the maze walker from discovering it.

\end{enumerate}


\subsection{Software Engineering Practices} 
\label{sub:software_engineering_practices}

A BPjs-based Java application is not a model, as it includes event handles, possibly a custom event selection strategy, and other mainstream programing constructs. However, when used properly, BPjs allows the core decision-making algorithms of said application to be a model. Thus, BPjs enables software engineers to verify (rather than test) core parts of the systems they develop.

A fully idiomatic b-program, where:
\begin{itemize}
  \item B-thread communicate through b-events only
  \item Reading inputs is done only through the external events queue
  \item Outputs are generated only through event listeners
\end{itemize}
is a model, since it can be formally analyzed, and, in particular, verified. 

\paragraph*{Iterative Process for Event Selection Strategies} As discussed in Subsection~\ref{subs:event-selection-strategies}, a good event selection strategy can greatly contribute to an application, both during the development process and at runtime. However, creating a good event selection strategy is not always trivial, especially during the first phases of a project. Building upon the hierarchical ordering of event selection strategies presented earlier in this section, we can propose an iterative development approach: First, start with a general strategy that might allow inefficient event selection. In subsequent iterations, refine the strategy. As long as the set of event traces possible under the refined strategy is a subset of the event traces possible under the basic strategy, the verified properties of the system will hold. Corollary: is it always possible to start with Simple strategy, and refine it as the project progresses. 

\section{Summary}
\label{sec:summary}

This paper presented consistent unified semantics for BP. Proposed semantics support almost all the existing body of work done in BP. Such semantics were hitherto unavailable, which we believe hindered research in this area. Additionally, this paper introduced BPjs, a generic platform for developing software systems based on the proposed BP semantics.

BPjs is intended to be a ``blue collar'' BP platform, to borrow a phrase from James Gosling's introduction of Java~\cite{gosling1997feel}. While its creation required novel ideas and generalizations, its usage should be easy and down to earth. As such, it can be used to introduce students and practitioners to BP, and to model-driven design in general.

BPjs can be improved in many ways, which we leave to future work (by us and others). These include better performance, lower memory requirements, new event selection strategies, improved debugging and logging tools, and a stronger verifier. Porting parts of the substantial body of work already developed using BP is also a worthwhile undertaking.

By providing a common environment that is both extensible and easy to use, BPjs allows researchers and industry to share and re-use tools and code. Similar, maybe, to ROS\cite{quigleyros} in robotics or R\cite{r-lang} and Zelig~\cite{zelig:paper,zelig:manual} in statistics. We hope BPjs becomes a boring part of BP research, a piece of software that makes creating the exciting parts easier.

\section*{Acknowledgment}
This type of project relies on the generosity of companies that provide free, high-quality infrastructure for open source projects. We thank GitHub, Sonatype (Maven Central), Travis CI GmbH, Coveralls.io, Readthedocs.io, and javadoc.io, whose excellent services are used in the development of BPjs.

The authors would like to thank David Harel and Assaf Marron for discussions, guidance, and hosting a presentation of an early version of this work at the Weizmann Institute of Science. Aviran Saadon, Achiya Elyasaf, and Meytal Genah, for using BPjs for their research, providing critical input and bearing with us while we fix bugs they have found. Moshe Weinstock, for technical contributions during the early stages of the project. Atilla Szegedi, for crucial work regarding Rhino continuations.

\bibliographystyle{plainnat}
\bibliography{./bibliography}

\end{document}